\documentclass{PoS}
\usepackage{graphicx}

\PoS{PoS(LAT2005)093}

\title{The static quark potential in 2+1 flavour Domain Wall QCD from QCDOC }

\ShortTitle{The static quark potential in 2+1 flavour Domain Wall QCD from QCDOC }

\author{\speaker{Koichi Hashimoto}\\
        Institute for Theoretical Physics, Kanazawa University, 
        Kanazawa 920-1192, Japan \\ 
        Radiation Laboratory, RIKEN, Wako 351-0198, Japan\\
        E-mail: \email{koichi@hep.s.kanazawa-u.ac.jp}}

\author{Taku Izubuchi\\
        Institute for Theoretical Physics, Kanazawa University, 
        Kanazawa 920-1192, Japan \\ 
        RIKEN-BNL Research Center, Brookhaven National Laboratory, 
        Upton, New York 11973, USA\\
        E-mail: \email{izubuchi@hep.s.kanazawa-u.ac.jp}}
        
\author{Jun Noaki\\
        School of Physics and Astronomy, University of Southampton,  
        Southampton SO17 1ST, England\\
        E-mail: \email{noaki@phys.soton.ac.uk}}
\author{RBC and UKQCD Collaborations}

\abstract{
  We report our present status of  on-going project on the
  measurement of the static quark potential in 2+1 flavour domain wall
  QCD with various improved gauge actions and couplings.  
  Lattice spacing determined from Sommer scale on these ensembles
  are from 1.6 GeV to 2.0 GeV  for $16^3\times 32$  lattice 
  with fifth dimension size 8. 
  We also examine size of discretization error from scaling 
  of a pair of dimensionless quantities, $(r_0 m_{\pi})^2$ and $r_0 m_\rho$,
  and found small scaling violation.
}

\FullConference{XXIIIrd International Symposium on Lattice Field Theory\\
		 25-30 July 2005\\
		 Trinity College, Dublin, Ireland}

\begin{document}

\section{Introduction}

The only way to remove pending issues associated with systematic
errors due to quenching in lattice QCD is carrying out simulation with
dynamical quarks.  Among several ongoing projects generating 2+1
flavour dynamical gauge ensembles, a striking characteristics of 
that by the RBC-UKQCD Collaborations is in their usage of 
domain wall fermions (DWF) for the sea quarks. 
DWF breaks chiral symmetry only by a small amount and thus 
has least discretization error.  There are a lot of
extensive measurements planned on the ensemble in near future.

We report our calculation of the static quark potential and 
preliminary results of lattice spacing $a$ from Sommer 
scale $r_0\approx 0.5$ fm \cite{Sommer scale}  
on the RBC-UKQCD gauge ensembles.

\section{Calculation}
The static potential $V(\vec{r})$ between a quark 
and anti-quark pair at relative spatial displacement 
$\vec{r}$,  is calculated from the Wilson loop $\langle W(\vec{r},t) \rangle$: 
$\langle W(\vec{r},t) \rangle = C(\vec{r}) e^{-V(\vec{r})t} + {\rm (``excited~states")},$
with a normalisation $C(\vec{r}=\vec{0})=1$.

We calculate $\langle W(\vec{r},t) \rangle$ in lattice QCD with 
2+1 flavour domain wall fermion~\cite{DWF action} and
several gauge actions (Iwasaki~\cite{Iwasaki action}, 
DBW2~\cite{DBW2 action}, and 
more negative rectangle coefficient,  
$c_{1}$, actions) on $16^{3}\times 32$ lattice with fifth dimension size, 
$L_s$=8, and domain wall height, $M_5$=1.8. 
The reason for calculating several gauge actions was  
to search for actions producing smaller 
residual mass, $m_{\rm res}$, at fixed $L_{s}$ and 
lattice spacing $a$ \cite{RM}. 
In quenched simulations, 
changing $c_{1}$ caused a drastic reduction of $m_{\rm res}$: 
$m_{\rm res}^{\rm (Wilson)}\approx10 \times 
m_{\rm res}^{\rm (Iwasaki)}\approx 
100 \times m_{\rm res}^{\rm (DBW2)}$ 
at $a^{-1}\approx$ 2 GeV \cite{quench DWF}.
We use three degenerate up and down quark masses,
$m_{u,d}=0.01$, 0.02, 0.04, and 
the strange quark mass $m_{s}=0.04$ was our best pre-simulation 
estimate of the physical strange quark mass. 
Simulation parameters are tabled in Table \ref{tab:Iwasaki}, \ref{tab:DBW2} 
and \ref{tab:lower c1}.

We compare two independent analyses, 
both of which  implement APE smearing \cite {APE smearing} 
for spatial links to improve the signal/noise.

\begin{figure}
\begin{tabular}{cc}
\begin{minipage}{0.50\textwidth}
\includegraphics[scale=0.40,clip=true]{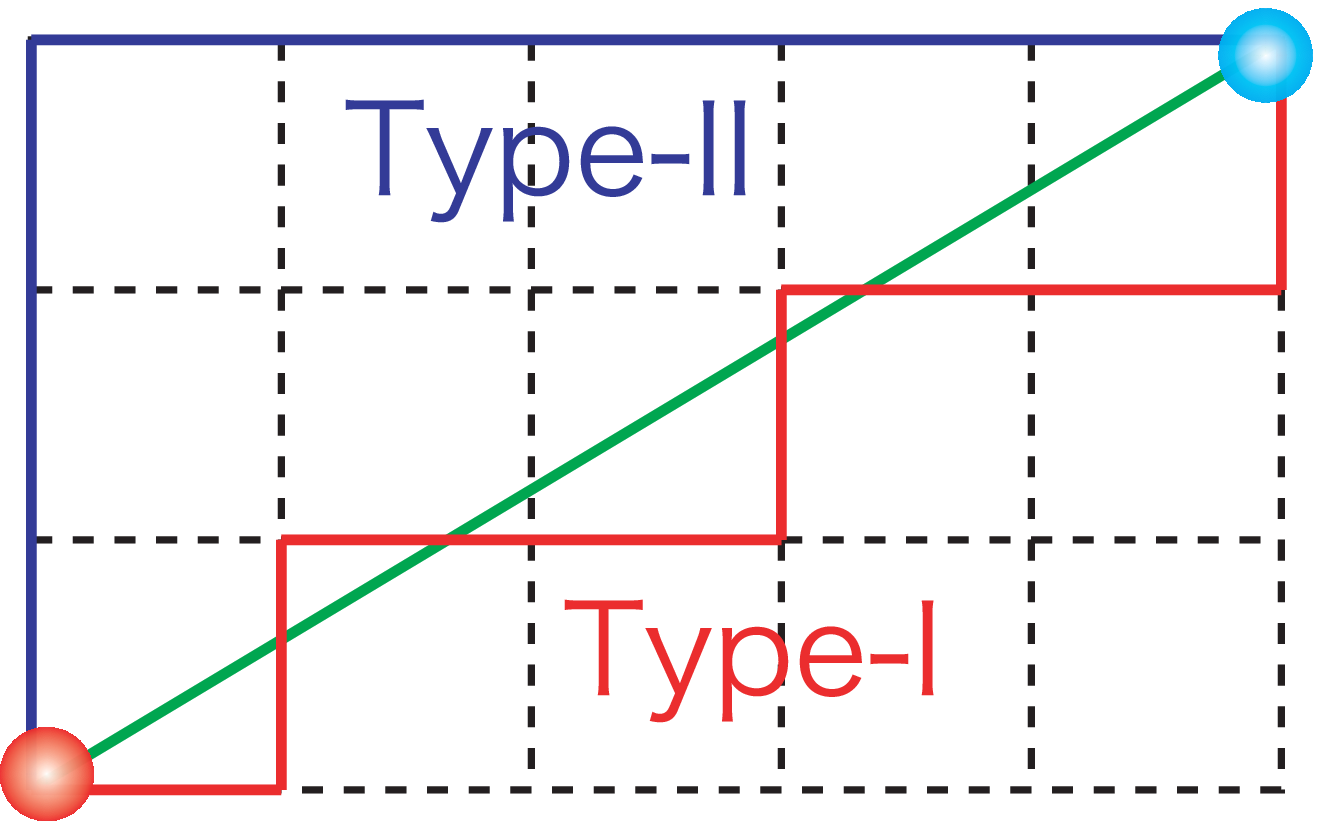}
\caption{Sketch of the path connecting quark and anti-quark. Type-I (red) is the closest path to diagonal line (green) by Bresenham algorithm, 
and Type-II (blue) is the taxi driver's path.}
\label{fig:path}
\end{minipage}

\begin{minipage}{0.50\textwidth}
\includegraphics[scale=0.40,clip=true]{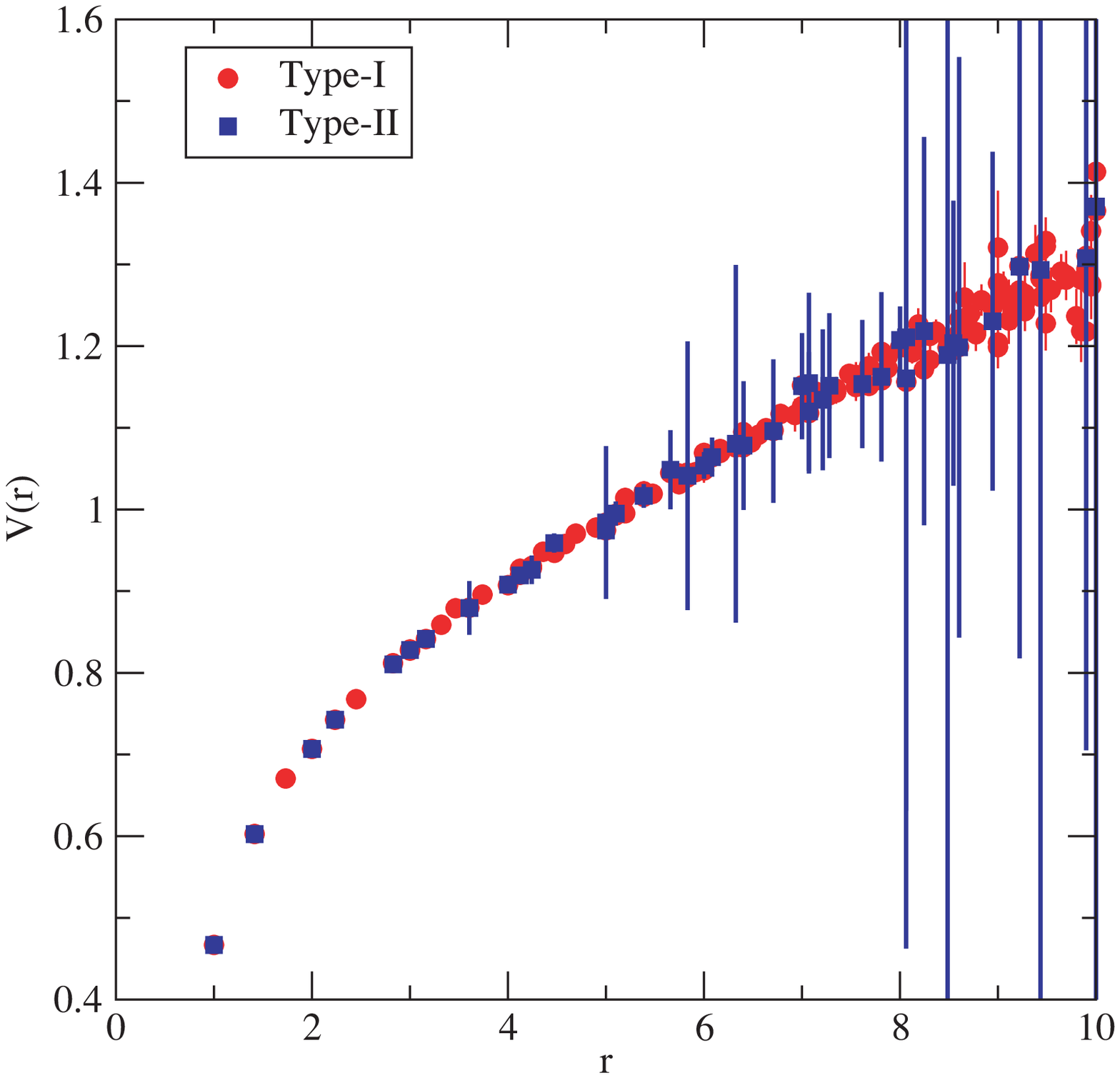}
\caption{$V(\vec{r})$ vs $|\vec{r}|$ for Iwasaki action, 
$\beta=2.20$, $m_{u,d}=m_{s}=0.04$.}
\label{fig:pot_diff_path}
\end{minipage}
\end{tabular}
\end{figure}

The main difference between these analyses is the path connecting 
the quark and anti-quark in a time slice 
(Figure \ref{fig:path}). 
The first analysis (Type-I) is carried out using the 
closest path to the diagonal line connecting the quark-antiquark. This path is 
determined 
by Bresenham algorithm~\cite{Bresenham algorithm} 
in the same way as \cite{dynamical DWF,KH}.
The second path (Type-II) is the taxi driver's path 
employed by Chroma code~\cite{Chroma}.
The smearing coefficient and number of smearing steps 
are tuned to be $(c_{\rm smear},N_{\rm smear})=(0.50, 20\sim40)$,
where the  overlap with ground state, $C(\vec{r})$, is approximately
its maximum.

The physical parameters are obtained 
by fitting the lattice value of $V(\vec{r})$ to the function:  
$V(\vec{r})=V_{0}-\alpha/ |\vec{r}|+\sigma |\vec{r}|,$
from which Sommer scale 
$r_{0}/ a=\sqrt{(1.65-\alpha) / \sigma}$ is determined in lattice unit.
For each gauge action and fixed gauge coupling $\beta$, 
we estimate lattice scale $a^{-1}$ assuming 
$r_{0}=0.5$ fm in the chiral limit.
Note this assumption turns out to be  consistent with the other 
scale setting ansatz using rho meson mass, $m_\rho=770$ MeV 
in previous simulations \cite{dynamical DWF,TI2}. 
\section{Results}

Figure \ref{fig:pot_diff_path} shows potentials measured 
using Type-I and Type-II paths.
These two analyses were performed on  same ensemble
($\beta=2.2$, $m_{u,d}=m_s=0.04$ on Table \ref{tab:Iwasaki}), 
and  central values are consistent with each other within statistical error. 
Sharply impoved signals (red circles) for relatively long distance 
off-axis spatial paths indicate that the $\bar Q$-${Q}$ state by Type-I 
has larger overlap with the ground state than that by Type-II. 
This is reasonably consistent with the picture of
the QCD-string connecting quark and anti-quark.

The time separation of Wilson loop, $t$, is chosen 
to minimise excited states contamination to the ground state
potential by monitoring $t$ dependence of $V(\vec{r})$. 
Then $V(\vec{r})$ is fit to the formula for $|\vec{r}| \in [r_{min}, r_{max}]$.
Figure \ref{fig:r0  Iwasaki DBW2} 
show $r_{0}$ as a function of dynamical quark mass, 
$m_{u,d}+m_{\rm res}$, for several fixed $\beta$'s. 
To estimate lattice spacing $a$,  we take the extrapolation 
to the point $m_{u,d}+m_{\rm res} = 0$ linearly and 
obtain $r_0/a$ in this limit (open symbols) from simulation points 
(filled symbols).
Preliminary results of this extrapolation and the value of $a^{-1}$
are summarised 
in Table \ref{tab:Iwasaki}, \ref{tab:DBW2} and \ref{tab:lower c1},
in which data at the simulation points are listed as well.
For the Type-I data, we estimate systematic errors by the dependence 
of the results on the fit ranges in extracting potentials from 
Wilson loop and in fitting the potentials to the specific function 
introduced in Section 1.
We estimate these errors from the variation of central values 
for $r_{min}\in[\sqrt{2},\sqrt{6}]$, ~$r_{max}\in[7,9]$, ~$t=5,6$.
Available results for Iwasaki $\beta=2.13$ and DBW2 $\beta=0.764$ 
are obtained by Type-II path. Analysis by Type-I path is under progress.
In particular, RBC-UKQCD has started their major production run using
Iwasaki $\beta=2.13$ on $24^3\times 64$ lattice with $L_{s}=16$. Our results
for Iwasaki $\beta=2.13$ provide properties of this ensemble.

\begin{figure}
\includegraphics[scale=0.40,clip=true]{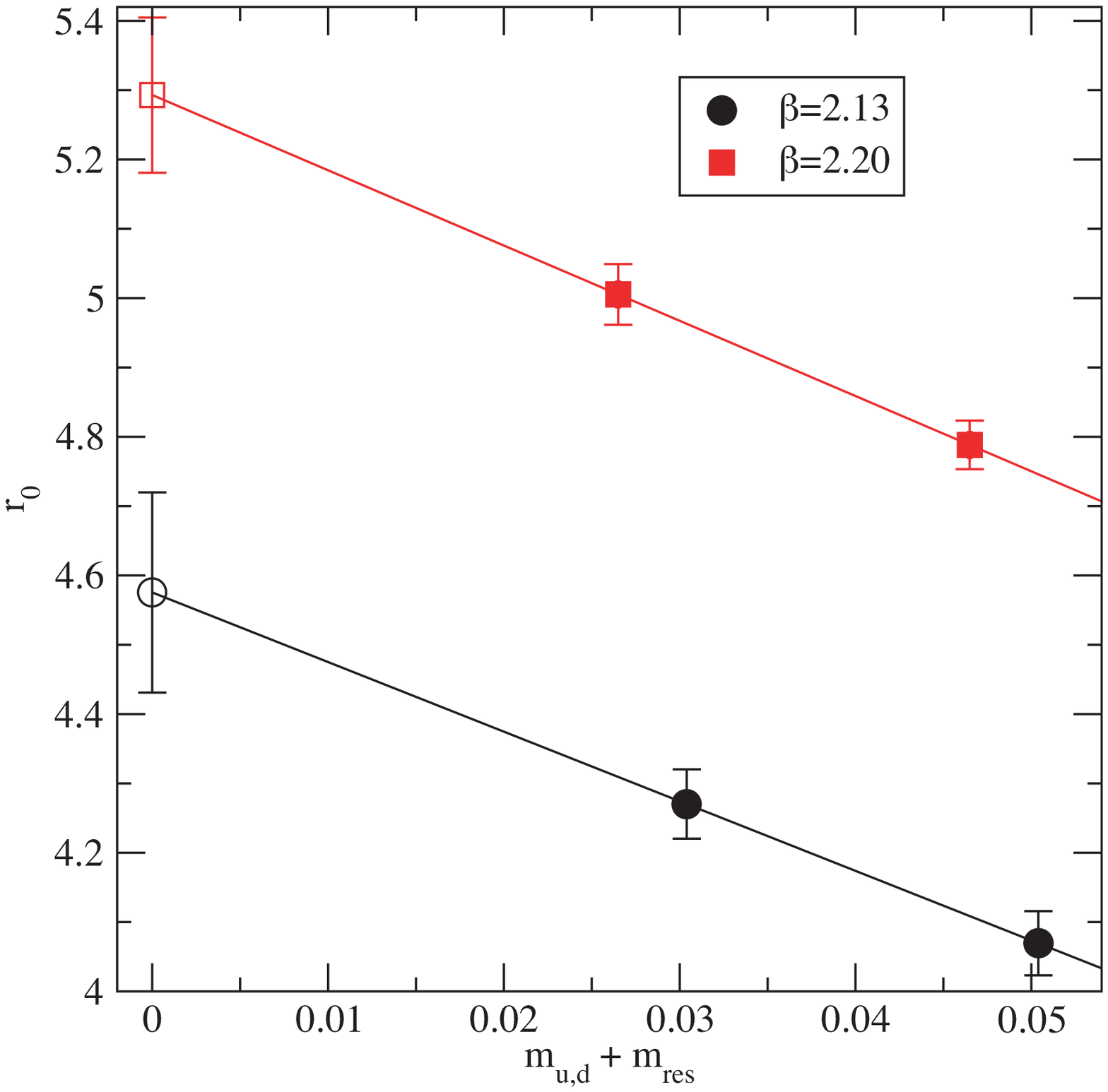}
\includegraphics[scale=0.40,clip=true]{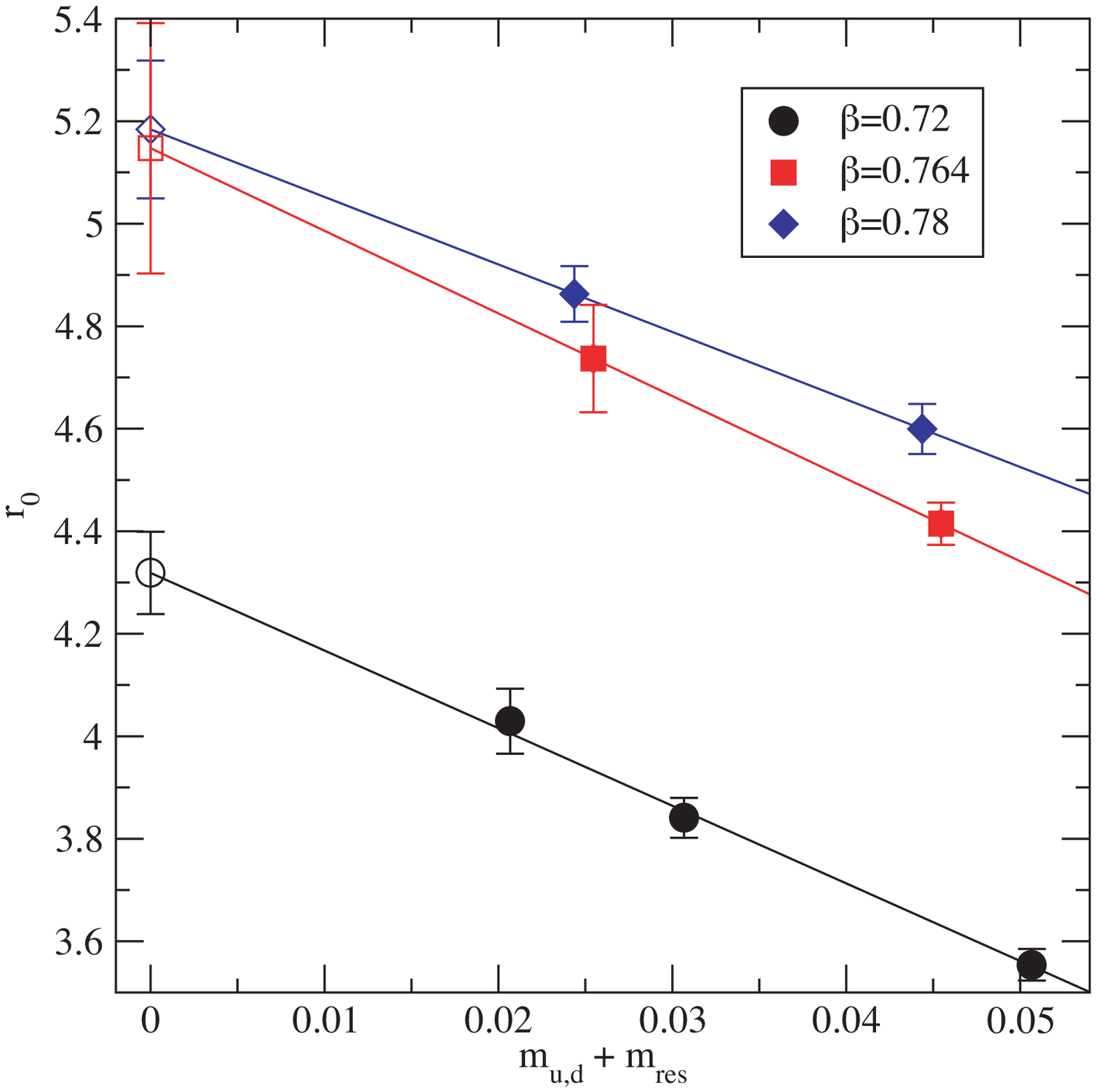}
\caption{
$r_{0}$ as a function of dynamical quark mass 
for  Iwasaki (left) and DBW2 (right) gauge action.
Each line shows linear extrapolation to the chiral limit, 
$m_{u,d} +m_{\rm res}\to 0$, for fixed $\beta$'s indicated in legends.
}
\label{fig:r0 Iwasaki DBW2}
\end{figure}

\begin{table}
\begin{tabular}{|c|c|c|c|c|c|c|c|}
\hline
$\beta$ & $m_{u,d}$ & \# conf & Alg & Path &
$r_0/a$ & $a^{-1}$ [GeV]\cr \hline 
  & 0.04 & 320 & 
  &  & 4.788(35)(30) & 1.890(14)(12) \cr 
 2.20 & 0.02 & 177 &
 RHMC & Type-I & 5.005(44)(48) & 1.975(17)(19) \cr
  & $-m_{\rm res}=-0.0065(1)$ & $-$ & 
  & & 5.293(112)(81) & 2.089(44)(32) \cr \hline
  & 0.04 & 306 & 
  & & 4.069(46) & 1.606(18) \cr 
 2.13 & 0.02 & 300 & 
 RHMC & Type-II & 4.270(50) & 1.685(20) \cr 
  & $-m_{\rm res}=-0.0104(2)$ & $-$ & 
  & & 4.576(145) &1.806(57) \cr \hline
\end{tabular}
\caption{
Results of $r_{0}$ for Iwasaki gauge action ($c_{1}=-0.331$).
$m_{s}=0.04$, $L_{s}=8$, $M_{5}=1.8$. 
Note our estimation for lattice cut off, $a^{-1}$, is the values
at chiral limit, $m_{u,d}\to-m_{\rm res}$.
}
\label{tab:Iwasaki}
\end{table}

\begin{table}
\begin{tabular}{|c|c|c|c|c|c|c|}
\hline
$\beta$ & $m_{u,d}$ & \# conf & Alg & Path &  
$r_0/a$ & $a^{-1}$ [GeV] \cr \hline
 0.80 & 0.04 & 200 & 
 Ralg & Type & 5.014(38)(67) & 1.979(15)(26) \cr 
  & & & & -I & & \cr \hline
  & 0.04 & 165 & 
  &  & 4.600(49)(16) & 1.815(19)(6) \cr 
 0.78 & 0.02 & 180 & 
 RHMC & Type & 4.863(54)(142) & 1.929(21)(56) \cr 
  & $-m_{\rm res}=-0.00437(6)$ & $-$ &
  & -I & 5.184(134)(335) & 2.046(53)(132) \cr \hline
  & 0.04 & 310 & 
  &  & 4.415(41) & 1.742(16) \cr 
 0.764 & 0.02 & 300 & 
 RHMC & Type & 4.737(105) & 1.869(41) \cr 
  & $-m_{\rm res}=-0.00546(7)$ & $-$ & 
  & -II & 5.147(244) & 2.031(96) \cr \hline
  & 0.04 & 280 & 
  & & 3.554(31)(67) & 1.403(12)(27) \cr 
 0.72 & 0.02 & 400 & 
 RHMC & Type &  3.841(39)(61) & 1.516(15)(24) \cr  
  & 0.01 & 150 & 
  & -I & 4.029(63)(27) & 1.590(25)(11) \cr  
  & $-m_{\rm res}=-0.01066(7)$ & $-$ & 
  & & 4.319(80)(28) & 1.704(32)(11) \cr \hline
\end{tabular}
\caption{Same as Table 1 for DBW2 action $(c_{1}=-1.4069)$. }
\label{tab:DBW2}
\end{table}

\begin{table}
\begin{tabular}{|c|c|c|c|c|c|c|}
\hline
$\beta$ & $c_{1}$ & \# conf & Alg & Path & 
$r_0/a$ & $a^{-1}$ [GeV] \cr \hline
 0.53 & $-2.3$ & 160 & 
 Ralg & Type-I & 5.074(48)(48) & 2.002(19)(19) \cr 
 0.48 & & 180 & 
 & & 4.002(57)(60) & 1.579(22)(24) \cr \hline 
0.36 & & 135 & 
 & & 5.060(66)(33) & 1.997(26)(13) \cr 
 0.33 & $-3.57$ & 145 & 
 Ralg & Type-I & 4.335(47)(102) & 1.711(19)(40) \cr 
 0.32 & & 180 & 
 & & 3.968(49)(43) & 1.566(19)(17) \cr \hline 
 0.16 & $-7.47$ & 165 & 
 Ralg & Type-I & 4.050(57)(43) & 1.598(22)(17) \cr \hline
\end{tabular}
\caption{
Results of $r_{0}/a$ and corresponding lattice scale $a^{-1}$
from the simulations with various $c_1$'s with $m_{u,d}= m_s=0.04$. 
}
\label{tab:lower c1}
\end{table}

\begin{figure}
\begin{center}
\includegraphics[scale=0.40,clip=true]{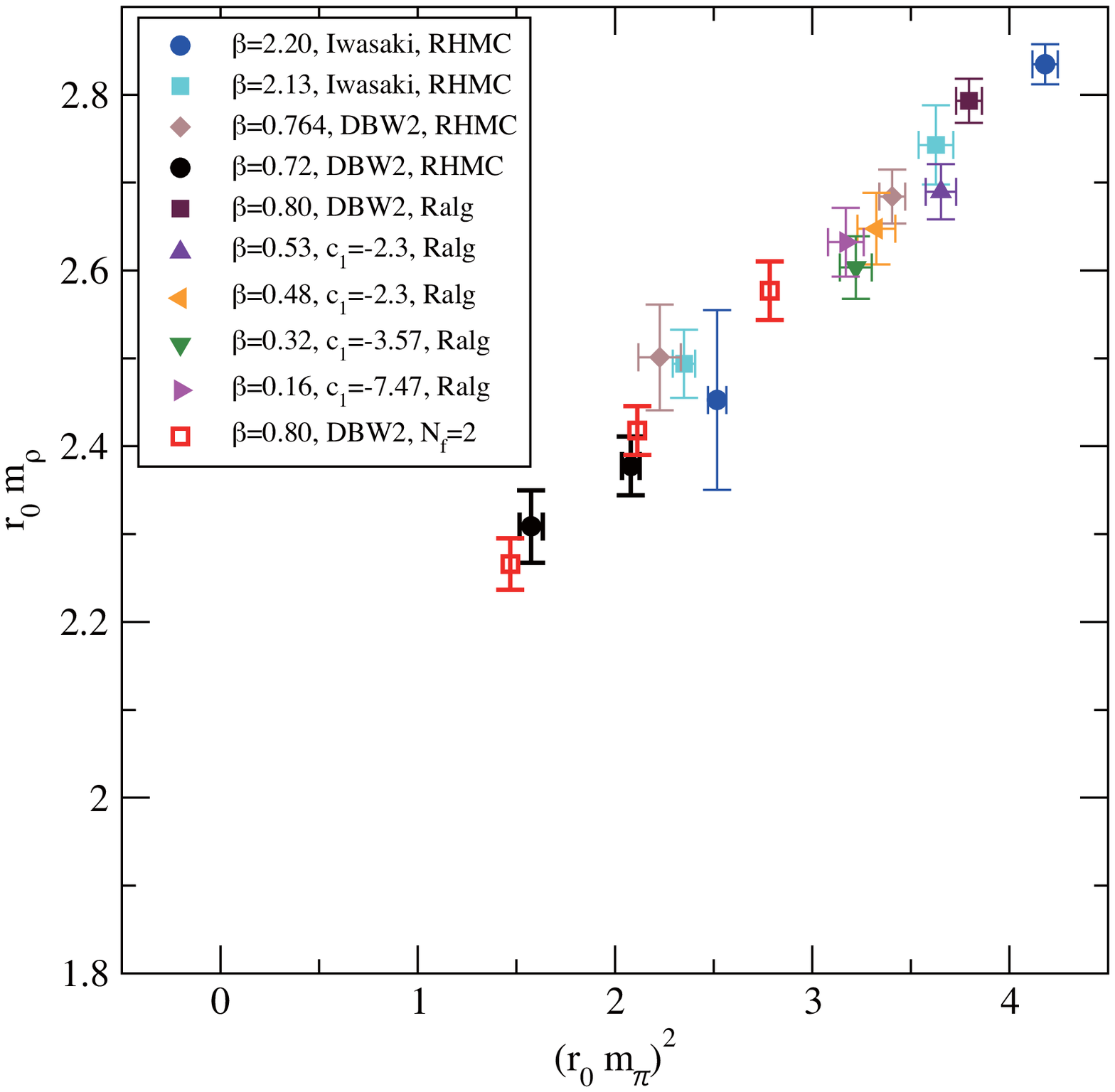}
\includegraphics[scale=0.40,clip=true]{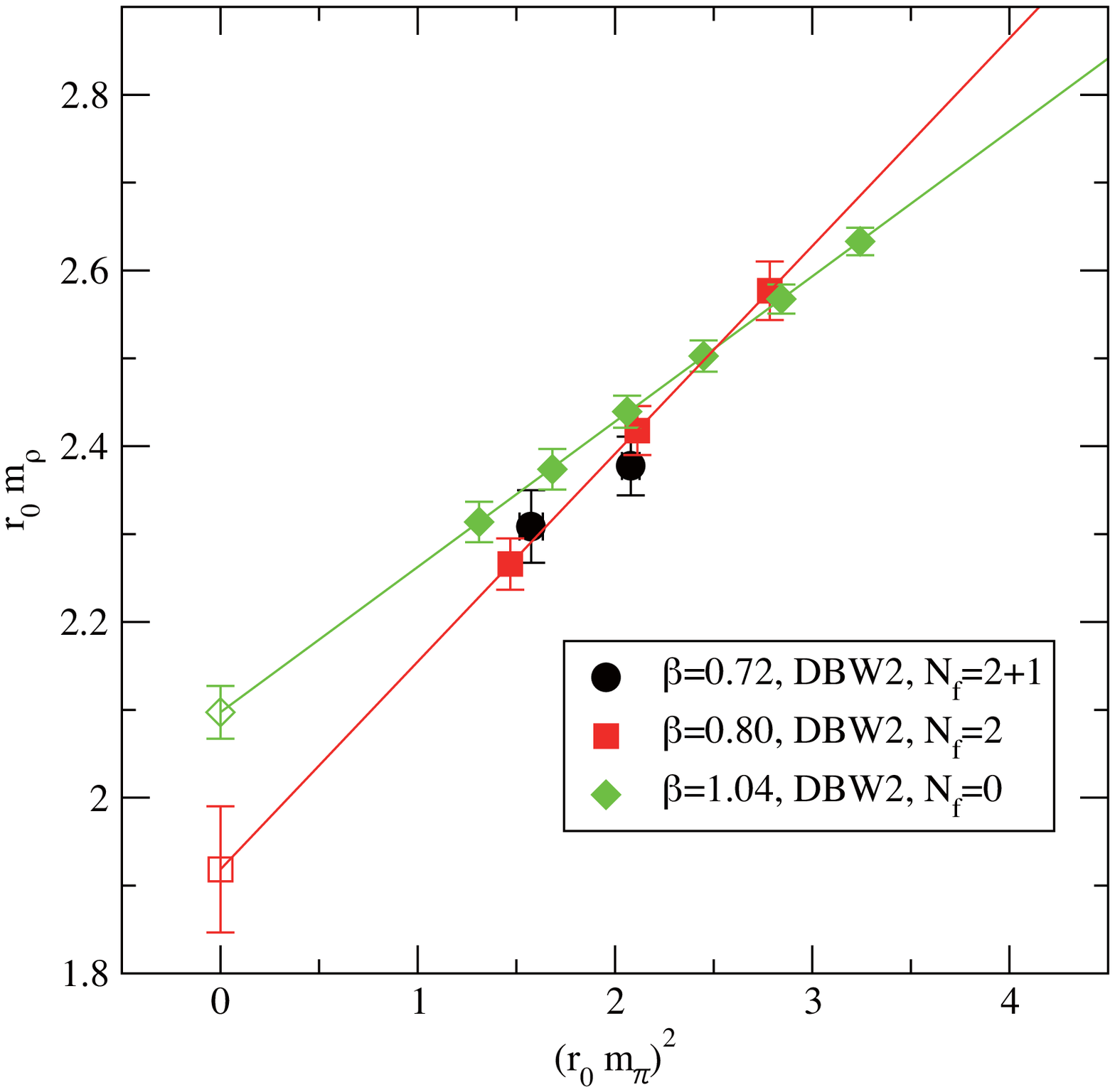}
\end{center}
\caption{
$r_0 m_\rho$ vs. $(r_0 m_\pi)^2$ for $N_f=2+1$ with $m_s=0.04$ (left). 
Right panel is the same plot for DBW2 $N_f=0$ $(\beta=1.04)$, 
$N_f=2$ $(\beta=0.80)$, 
and $N_f=2+1$ $(\beta=0.72,~m_s=0.04)$ with linear chiral extrapolation
$(r_0 m_\pi)^2 \to 0$ for $N_f=2$ and 0.
Error is uncorrelated between $r_{0}$ and meson spectra.}
\label{fig:r0_times_mrho}
\end{figure}

Finally we measure the magnitude of discretization error using the data 
from all ensembles we measured. Due to the (almost exact) chiral symmetry 
of DWF,  
the scaling violation is expected  as
${\cal O}( (a\Lambda_{QCD})^2)
+{\cal O}(a m_{\rm res})
\sim {\cal O}(1)$\% 
for the parameters used. We examine this expectation by plotting a pair of
dimensionless quantities, 
$((r_{0} m_{\pi})^{2}, r_{0} m_{\rho})$ in Figure \ref{fig:r0_times_mrho}. 
$m_\pi$, $m_\rho$ are  pseudo-scalar and vector meson masses
respectively  at relatively large valence quark mass set to be equal
to sea quark mass $m_{u,d}$ \cite{three}. 
If scaling were perfect (no discretization error), 
all data points are 
on a universal curve in the two dimensional plane. 
The left panel of Figure \ref{fig:r0_times_mrho} demonstrates 
the expectation is 
true for these quantities: points are on a line within statistical
\def\gsim{\raisebox{0.8mm}{\em$\,<$}\hspace{-2.45mm}\raisebox{-0.9mm}{\em $\sim \,$}}
error ($\gsim$ 5 \% ). 

We also compare DBW2 $\beta=0.80$  results with those of two flavour
(red square) and quenched (green diamonds) simulation in the right panel. 
The universal line of two flavour points roughly accommodate the 2+1
flavour points. It is expected because $m_s$ is heavy and has less 
effect on up and down quark meson masses than $u$ and $d$ quarks.
On the other hand, quenched data have significantly smaller slope than 
other two data sets leading larger value of $r_0 m_\rho$ in the chiral limit.

However, we note that the estimation of statistical errors are not so accurate 
that they reflect the actuality completely since the length of total 
trajectory and the separation between measured trajectories may not be large 
enough for some of ensembles.

\section*{Acknowledgement}
We thank Peter Boyle, Mike Clark, Saul Cohen, 
Sam Li, Meifeng Lin, Chris Maynard, and Robert Tweedie, and Azusa Yamaguchi
for generating ensembles and meson masses used in this work. 
We thank Peter Boyle, Dong Chen, Norman Christ, Mike Clark,
Saul Cohen, Calin Cristian, Zhihua Dong, Alan Gara, Andrew Jackson, Balint Joo,
Chulwoo Jung, Richard Kenway, Changhoan Kim, Ludmila Levkova, Xiaodong Liao,
Guofeng Liu, Robert Mawhinney, Shigemi Ohta, Konstantin Petrov, Tilo Wettig and Azusa Yamaguchi
for developing with us the QCDOC machine and its software. This development
and the resulting computer equipment used in this calculation were funded
by the U.S. DOE grant DE-FG02-92ER40699, PPARC JIF grant PPA/J/S/1998/00756
and by RIKEN. This work was supported by DOE grant DE-FG02-92ER40699 and
we thank RIKEN, BNL, the U.S. DOE and University of Edinburgh 
for providing the facilities essential
for the completion of this work.
We also thank the RIKEN Super Combined Cluster (RSCC) at RIKEN, for 
the computer resources used for the static quark potential calculation. 
K.H. thanks RIKEN BNL Research Center for its hospitality where this work 
was performed.


\begin{thebibliography}{99}

\bibitem{Sommer scale}
R. Sommer, 
{\it ``A New Way to Set the Energy Scale in Lattice Gauge Theories and its Application to the Static Force and $\alpha_s$ in SU(2) Yang--Mills Theory''}, 
Nucl. Phys. B411 (1994) 839, {[\tt hep-lat/9310022]}.

\bibitem{DWF action}
D. B. Kaplan, 
{\it ``A Method for Simulating Chiral Fermions on the Lattice''}, 
Phys. Lett. B288 (1992) 342, {[\tt hep-lat/9206013]}; 
Y. Shamir, 
{\it ``Chiral Fermions from Lattice Boundaries''}, 
Nucl. Phys. B406 (1993) 90, {[\tt hep-lat/9303005]}; 
V. Furman, Y. Shamir, 
{\it ``Axial symmetries in lattice QCD with Kaplan fermions''}, 
Nucl. Phys. B439 (1995) 54 {[\tt hep-lat/9405004]}.

\bibitem{Iwasaki action}
Y. Iwasaki, 
{\it ``Renormalization Group Analysis of Lattice Theories and Improved Lattice Action''}, 
uTHEP-117, uTHEP-118 (1983); 
Y. Iwasaki and T. Yoshie, 
{\it ``Renormalization group improved action for $SU(3)$ lattice gauge theory and the string tension''}, 
Phys. Lett. B143 (1984) 449.

\bibitem{DBW2 action}
T. Takaishi, 
{\it ``Heavy quark potential and effective actions on blocked configurations''}, 
Phys. Rev. D54 (1996) 1050; 
P. de Forcrand {\it et al.} (QCD-TARO), 
{\it ``Renormalization group flow of SU(3) lattice gauge theory - Numerical studies in a two coupling space''}, 
Nucl. Phys. B577 (2000) 263 {[\tt hep-lat/9911033]}.

\bibitem{RM}
R. D. Mawhinney (RBC and UKQCD) in these proceedings.

\bibitem{quench DWF} 
Y. Aoki {\it et al.} (RBC), 
{\it ``Domain wall fermions with improved gauge actions''}, 
Phys. Rev. D69 (2004) 074504, {\tt [hep/lat-0211023]}.


\bibitem{APE smearing}
M. Albanese {\it et al.} (APE), 
{\it ``Glueball masses and String tension in Lattice QCD''}, 
Phys. Lett. B192 (1987) 163. 

\bibitem{Bresenham algorithm}
B. Bolder {\it et al.}, 
{\it ``A High Precision Study of the $Q\bar Q$ Potential from Wilson Loops in the Regime of String Breaking''}, 
Phys. Rev. D63 (2001) 074504, {\tt [hep-lat/0411006]}.

\bibitem{dynamical DWF}
Y. Aoki {\it et al.} (RBC), 
{\it ``Lattice QCD with two dynamical flavors of domain wall quarks''}, 
{\tt hep-lat/0411006}.

\bibitem{KH}
K. Hashimoto and T. Izubuchi (RBC), 
{\it ``Static $\bar Q$-$Q$ Potential from $N_{f}=2$ Dynamical Domain-Wall QCD''}, 
Nucl. Phys B140 (Proc. Suppl.) (2005) 341, {\tt [hep-lat/0409101]}.

\bibitem{Chroma}
R. G. Edwards and B. Joo, 
{\it ``The Chroma Software System for Lattice QCD''}, 
Nucl. Phys B140 (Proc. Suppl.) (2005) 832, {\tt [hep-lat/0409003]}.

\bibitem{TI2}
T. Izubuchi (RBC), 
{\it ``Hadron Spectrum and Decay Constant from $N_F=2$ Domain Wall QCD''}, 
Nucl. Phys B140 (Proc. Suppl.) (2005) 237, {\tt [hep-lat/0410034]}.


\bibitem{three}
M. Lin (RBC and UKQCD) in these proceedings; 
C. M. Maynard (RBC and UKQCD) in these proceedings; 
R. Tweedie (RBC and UKQCD) in these proceedings.


\end{thebibliography}
\end{document}